\documentclass[aip, apl, onecolumn, superscriptaddress, reprint]{revtex4-1}

\usepackage{graphicx}
\usepackage{bm}
\usepackage{dcolumn}
\usepackage{amssymb,amsmath}
\usepackage[colorlinks=true,allcolors=blue]{hyperref}
\usepackage{color}
\usepackage{siunitx}

\usepackage{lipsum}
\newcommand{\cOut}[1]{}
\newcommand{\figref}[2]{\hyperref[#1]{\ref{#1}(#2)}}

\begin{document}

\title{A strain-controlled magnetostrictive pseudo spin valve}



\author{Vadym Iurchuk}
\thanks{Current affiliation: Institute of Ion Beam Physics and Materials Research, Helmholtz-Zentrum Dresden-Rossendorf, 01328 Dresden, Germany}
\email[e-mail: ]{v.iurchuk@hzdr.de}
\affiliation{Institut de Physique et Chimie des Matériaux de Strasbourg (IPCMS), UMR 7504 CNRS, Université de Strasbourg, 23 rue du Loess, 67034, Strasbourg, France}

\author{Julien Bran}
\affiliation{Institut de Physique et Chimie des Matériaux de Strasbourg (IPCMS), UMR 7504 CNRS, Université de Strasbourg, 23 rue du Loess, 67034, Strasbourg, France}

\author{Manuel Acosta}
\affiliation{Institut de Physique et Chimie des Matériaux de Strasbourg (IPCMS), UMR 7504 CNRS, Université de Strasbourg, 23 rue du Loess, 67034, Strasbourg, France}

\author{Bohdan Kundys}
\affiliation{Institut de Physique et Chimie des Matériaux de Strasbourg (IPCMS), UMR 7504 CNRS, Université de Strasbourg, 23 rue du Loess, 67034, Strasbourg, France}

\date{\today}

\begin{abstract}
    Electric-field control of magnetism via inverse magnetostrictive effect is an efficient path towards improving energy-efficient storage and sensing devices based on giant magnetoresistance effect. In this letter, we report on lateral electric-field driven strain-mediated modulation of magnetic properties in Co/Cu/Py pseudo spin valve grown on ferroelectric 0.7Pb[Mg$_{1/3}$Nb$_{2/3}$)]O$_3$–0.3PbTiO$_3$ substrate. We show a decrease of the giant magnetoresistance ratio of the pseudo spin valve with increasing electric field, which is attributed to the deviation of the Co layer magnetization from the initial direction due to strain-induced magnetoelastic anisotropy contribution. Additionally, we demonstrate that strain-induced magnetic anisotropy effectively shifts the switching field of the magnetostrictive Co layer, while keeping the switching field of the nearly zero-magnetostrictive Py layer unaffected due to its negligible magnetostriction constant. We argue that magnetostrictively optimized magnetic films in properly engineered multilayered structures can offer a path to enhancing the selective magnetic switching in spintronic devices.
\end{abstract}

\maketitle

Modern spintronics demands for high-stability, low-power, fast-switching devices operable at room temperature. Advanced approaches of magnetization control tend to promote alternative to magnetic field methods to get access to magnetic state of spintronic elements, e.g. magnetic field sensors or magnetoresistive memory cells~\cite{bibes_multiferroics:_2008, shen_multilevel_2016, vaz_electric_2012, binek_magnetoelectronics_2005}. One of the promising ways is electric-field control of magnetic properties in artificial ferroelectric/ferromagnetic heterostructures via strain-mediated magnetoelectric effects~\cite{domann_strain-mediated_2018}. Strain-controlled tuning of magnetism has become a hot topic of recent theoretical~\cite{roy_switching_2011, roy_hybrid_2011, roy_energy_2012, roy_ultra-low-energy_2013, jia_piezoelectric_2012, biswas_complete_2014, wang_strain-mediated_2017, wang_strain-mediated_2018, schneider_rf_2019} and experimental~\cite{bihler_textga_1ensuremath-xtextmn_xtextaspiezoelectric_2008, ranieri_lithographically_2008, rushforth_voltage_2008, brandlmaier_textitsitu_2008, brandlmaier_nonvolatile_2011, brandlmaier_magneto-optical_2012, liu_tunable_2011, zhang_anisotropic_2012, sando_crafting_2013, lahtinen_electric-field_2012, liu_non-volatile_2013, lei_strain-controlled_2013, buzzi_single_2013, zhu_effects_2013, peng_switching_2018, zhu_strain-driven_2021, iurchuk_multistate_2014, iurchuk_electrical_2015} studies, where a piezoelectric strain is shown to modify magnetic anisotropy, coercivity, anisotropic magnetoresistance (AMR) or switch the magnetization direction of the magnetic component of magnetoelectric heterostructures. From the practical point of view, it is important to utilize this effect in spintronic structures, e.g. in giant magnetoresistive (GMR) multilayers or magnetic tunnel junctions. Several successful attempts were done to confirm the viability of the strain-mediated manipulation of giant or tunneling magnetoresistance in magnetic multilayers grown on ferroelectric~\cite{li_electric_2014, roy_separating_2015, chen_giant_2019, liu_nonvolatile_2021} or mechanically flexible~\cite{liu_magnetostrictive_2016, hashimoto_spin-valve_2018} substrates. However, commonly used design of the magnetic devices with magnetostrictive functionality requires the presence of the pinning antiferromagnetic layer in the magnetic stack to ensure a separate access and control over the magnetic layers.

In this letter, we report on lateral electric-field induced changes of GMR properties in PMN-PT/Co/Cu/Py (Py=Ni$_{80}$Fe$_{20}$) heterostructure comprising magnetic layers with different magnetostriction constants. We show that the GMR ratio of the Co/Cu/Py pseudo spin valve (PSV) as well as the switching field of the Co layer decrease as electric-field induced piezoelectric strain increases. We attribute this effect to the strain-induced modification of the magnetic anisotropy of the Co layer through the inverse magnetostrictive (magnetoelastic) effect. On the other hand, the switching field of the Py layer changes only scarcely under strain, which is a consequence of known close-to-zero magnetostriction of the Ni$_{80}$Fe$_{20}$ Permalloy compound. This result paves a way to strain-controlled manipulation of magnetic and magnetotransport properties of PSV tri-layers with properly adjusted magnetostrictive properties of the constituent magnetic layers.

\begin{figure*}[ht!]
    \centering
    \includegraphics[width=0.95\textwidth]{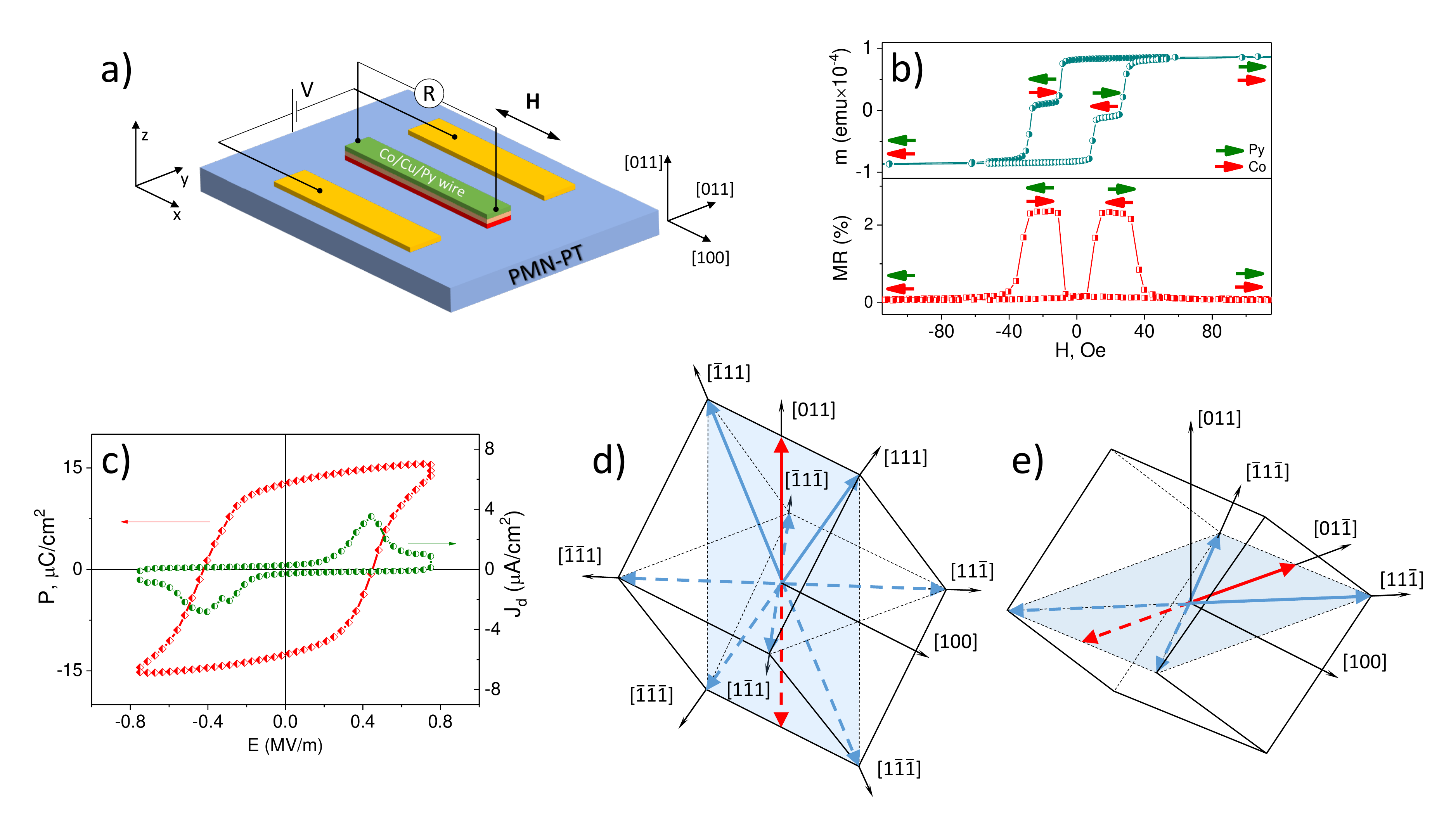}
    \caption{(a) Schematics of the Co/Cu/Py pseudo spin valve stripe fabricated on PMN–PT single crystalline substrate. (b) Magnetic and magnetoresistance hysteresis loops measured on Si/SiO$_2$/Co(5)/Cu(4.5)/Py(5) (thicknesses is nm) (same deposition batch with PMN-PT/Co/Cu/Py) showing distinct and well-separated switching fields of Py (green arrow) and Co (red arrow) layers. (c) Ferroelectric polarization and corresponding displacement current of PMN-PT single crystal measured in the lateral geometry as shown in (a). (d, e) Schematic representation of the distorted PMN–PT rhombohedral crystal structure with highlighted (blue arrows) polarization directions in out-of-plane (d) and in-plane (e) ferroelectric domains. Red arrows denote net polarization directions. Lattice distortions are exaggerated.}
    \label{Fig1}
\end{figure*}

The composition of the PSV tri-layer was chosen as Co/Cu/Py, motivated by relatively large magnetostriction constant of Co --20 to --60 ppm~\cite{klokholm_saturation_1982} depending on the crystallographic phase admixture in the sputtered polycrystalline films) and negligible magnetostriction constant of nearly zero-magnetostrictive Py (notably its Ni$_{80}$Fe$_{20}$ compound with $\lambda_s \approx$ 1.5 ppm)~\cite{reekstin_zero_1967, klokholm_saturation_1981}. The Co/Cu/Py PSV was deposited onto 500 µm thick polished ferroelectric 0.7Pb[Mg$_{1/3}$Nb$_{2/3}$)]O$_3$–0.3PbTiO$_3$ (PMN–PT) single crystalline substrate by magnetron sputtering at a base pressure of $ 3 \times 10^{-7} $ mbar. A shadow mask was used to obtain patterned $\sim$950 $\times$ 35 $\mu$m$^2$ rectangular stripes. No magnetic field was applied during the deposition. Ti(5 nm)/Au(50 nm) surface electrodes were deposited aside the stripe (see Fig.~\ref{Fig1}(a)) to generate piezoelectric strain as a result of the lateral electric field stimulus of the ferroelectric substrate. Two-probe magnetoresistance (MR) measurements were carried out in current-in-plane geometry with magnetic field applied parallel to the long axis of the PSV stripe. Fig.~\ref{Fig1}(a) shows the schematics of the sample under study and the geometry of the experiment.
 
A set of polycrystalline tri-layers was prepared varying the thicknesses of the magnetic and the spacing layers. The optimal composition for the presented study was chosen to be Co(5)/Cu(4.5)/Py(5) (thicknesses are in nm) since it provides both significant GMR ratio and clear separation between the coercive fields of Co and Py layers. Fig.~\ref{Fig1}(b, top graph) shows the room-temperature magnetic hysteresis of the PSV tri-layer performed on the reference sample (grown on Si/SiO$_2$ substrate in the same deposition batch with PMN-PT). We observe clear and relatively sharp switching of both layers forming a well-defined plateau between the two coercive fields corresponding to the magnetization reversal of Py (red arrow) and Co (green arrow) layers respectively. The MR loop measured on the extended Co/Cu/Py tri-layer on Si/SiO$_2$ perfectly correlates with the magnetic hysteresis (see Fig.~\ref{Fig1}(b, bottom graph)), yielding a GMR ratio of 2.3$\%$ at room temperature.

An unpoled rhombohedral (011)PMN–PT ferroelectric single crystal (commercially available from \textit{Crystal GmbH}) was used as a functional substrate to generate an electric-field-induced strain, which is transferred to the PSV tri-layer grown atop. The PMN–PT composition is in the vicinity of the ferroelectric morphotropic phase boundary~\cite{guo_effect_2002} where large piezoelectric coefficients~\cite{zhang_elastic_2008} are expected allowing for large strains when moderate electric fields are applied to the crystal. Fig.~\ref{Fig1}(c) shows the ferroelectric polarization and the electrical displacement current loops vs. applied electric field measured in the lateral geometry (see Fig.~\ref{Fig1}(a)). Ferroelectric coercive field for this particular geometry is $E_C$ = 0.43 MV/m which is approximately twice higher comparing to the typical values for the out-of-plane geometry, when the electric field is applied across the bulk of the substrate~\cite{zhang_elastic_2008, wu_domain_2011}. The origin of this discrepancy is likely attributed to the sub-surface ferroelectric domains growth and reversal, where a net polarization switches only in topmost thin PMN–PT layer while the bulk ferroelectric state remains unchanged.

Ferroelectric polarization reversal is accompanied by ferroelastic domains switching which leads to the distortions of the crystal lattice. Schematic images of polarization orientations in PMN–PT crystal with rhombohedral structure are presented in Fig.~\ref{Fig1}(d,e). When no electric field is applied the spontaneous ferroelectric polarization is aligned along one of the eight equivalent $<$111$>$ directions (blue arrows), creating ferroelectric domains with both out-of-plane and in-plane net polarization (red arrows). Application of a transverse electric field (here in [01$\bar{1}$] direction) forces the polarization of out-of-plane domains to switch along the direction of the applied electric field. In our case, the electric field is applied along [01$\bar{1}$] crystal axis resulting in the sizable lattice deformation during the ferroelastic transition which generates a tensile strain in this direction (Fig.~\ref{Fig1}(e)). The strain is transferred through the ferroelectric/ferromagnetic interface to the PSV structure and induces modifications of its static magnetic properties due to the inverse magnetostrictive (magnetoelastic) effect.

\begin{figure}[t]
    \centering
    \includegraphics[width=7cm]{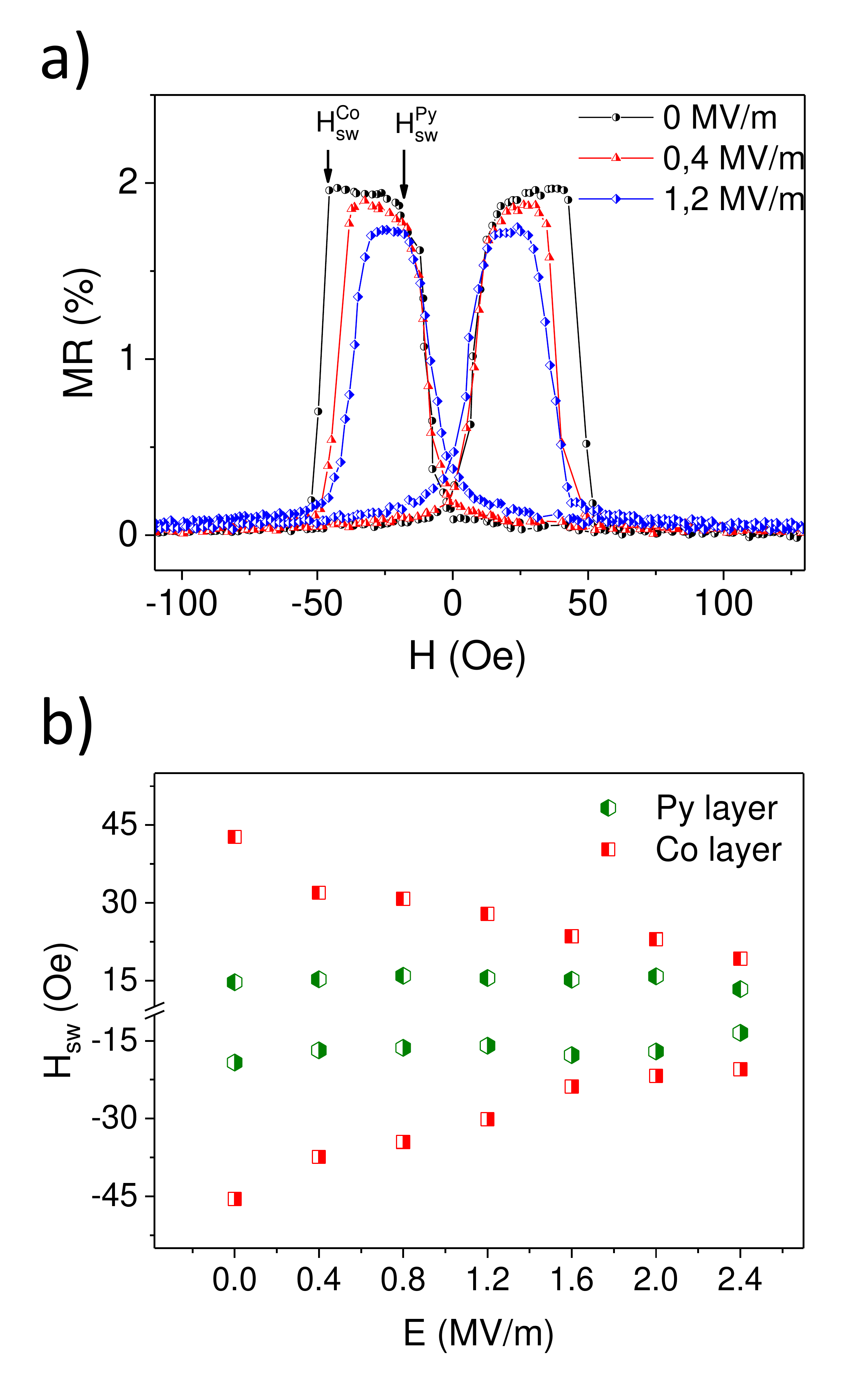}
    \caption{(a) Magnetoresistance loops of Co(5)/Cu(4.5)/Py(5) PSV stripe on PMN–PT measured for different lateral electric fields applied to the substrate. (b) Switching fields of Co (red squares) and Py (green circles) layers extracted from (a).}
    \label{Fig2}
\end{figure}

The MR loops of the Co/Cu/Py PSV tri-layer are investigated for the magnetic field $H$ applied parallel to the probing current $I$ and under different electric fields applied to the ferroelectric PMN–PT substrate (Fig.~\ref{Fig2}(a)). At zero electric field (black graph), MR value reaches $\sim$2 $\%$, and the shape of the loop indicates the sharp and separate switching of both Py and Co magnetic layers with the switching fields $H_{sw}^{Py} \approx $ 15 Oe and $H_{sw}^{Co} \approx $ 45 Oe respectively. Here, we define the switching field as the onset magnetic field where the peak MR value at the plateau starts decreasing from its maximum value. These values are in good agreement with the switching process in the reference sample grown on Si/SiO$_2$ substrate (see Fig.~\ref{Fig2}(b)). Slight deviations in the MR ratio and the switching field values may arise due to small interface strains during the deposition or different PMN–PT surface roughness as compared to Si/SiO$_2$. When the lateral electric field is applied to the substrate, the shape of the MR curves is notably modified (see red and blue graphs in Fig. 2(a)). Upon increasing electric field, we can distinct three superimposed effects: i) narrowing of the plateau between the switching fields $H_{sw}^{Py}$ and $H_{sw}^{Py}$; ii) increase of the MR peak steepness and iii) decrease of the GMR ratio of the PSV. The observed narrowing of the plateau between the switching fields of the magnetic layers comparing to the initial state originates from the decrease of the Co layer switching field $H_{sw}^{Co}$. Fig.~\ref{Fig2}(b) shows both positive and negative values of the switching fields of Co and Py magnetic layers vs. applied electric field extracted from the corresponding MR loops. There is only marginal impact of the electric field on the switching field of the Py layer, whereas a monotonous decrease of the switching field of the Co layer is observed when increasing the electric field. A gradual increase of the MR peak steepness with increasing electric field (see Fig.~\ref{Fig2}(a)) indicates the magnetic hardening of the $x$-axis of the stripe. 
 
The described effects are explained as a result of the strain-dependent magnetic anisotropy in ferroelectric/ferromagnetic heterostructures. When a lateral electric field is applied to the ferroelectric crystal in a spontaneous ferroelectric state, the converse piezoelectric effect results in an in-plane strain (see Fig.~\ref{Fig1}(e)). This strain, when transferred to the PSV tri-layer, modifies the magnetic anisotropies of the magnetic layers via the inverse magnetostrictive (magnetoelastic) effect. Both Co and Py magnetic layers acquire an additional strain-dependent magnetoelastic energy $E_{\sigma} = -\frac{3}{2} \lambda_s Y \epsilon cos^2 \theta$ , where $\lambda_s$ is the magnetostriction constant, $Y$ is the Young’s modulus of the magnetic material, $\epsilon$ is the generated stress and $\theta$ is the angle between the magnetization and strain directions. Given the close-to-zero value of the magnetostriction constant of the Py compound~\cite{reekstin_zero_1967}, the corresponding magnetoelastic contribution under 
strain is marginal and, therefore, insufficient to induce any noticeable change in the magnetic anisotropy of the Py layer. Therefore, the switching field $H_{sw}^{Py}$ of the Py layer does not change significantly with increasing the electrically induced strain in the structure. On the other hand, the inverse magnetostrictive effect introduces a sizable magnetoelastic energy contribution to the magnetic anisotropy of the Co layer, which depends on the value of the electric-field-induced strain and on the magnetostriction constant of Co. Since polycrystalline Co thin films have negative magnetostriction constant~\cite{klokholm_saturation_1982}, the tensile strain along y-axis (see Fig.~\ref{Fig1}(a)) introduces an additional magnetoelastic anisotropy along the transverse direction (here $x$-direction), making the $x$-axis more preferable for the magnetization arrangement. This manifests in the decrease of the switching field $H_{sw}^{Co}$ along the $x$-axis of the Co layer. This explanation is also consistent with the gradual change of the shape of the MR curves upon increasing strain, from the square-shaped easy-axis loop at 0 MV/m to the triangular-shaped hard-axis loop at 2.4 MV/m (see Fig~\ref{Fig2}(a)).

\begin{figure}[h]
    \centering
    \includegraphics[width=7cm]{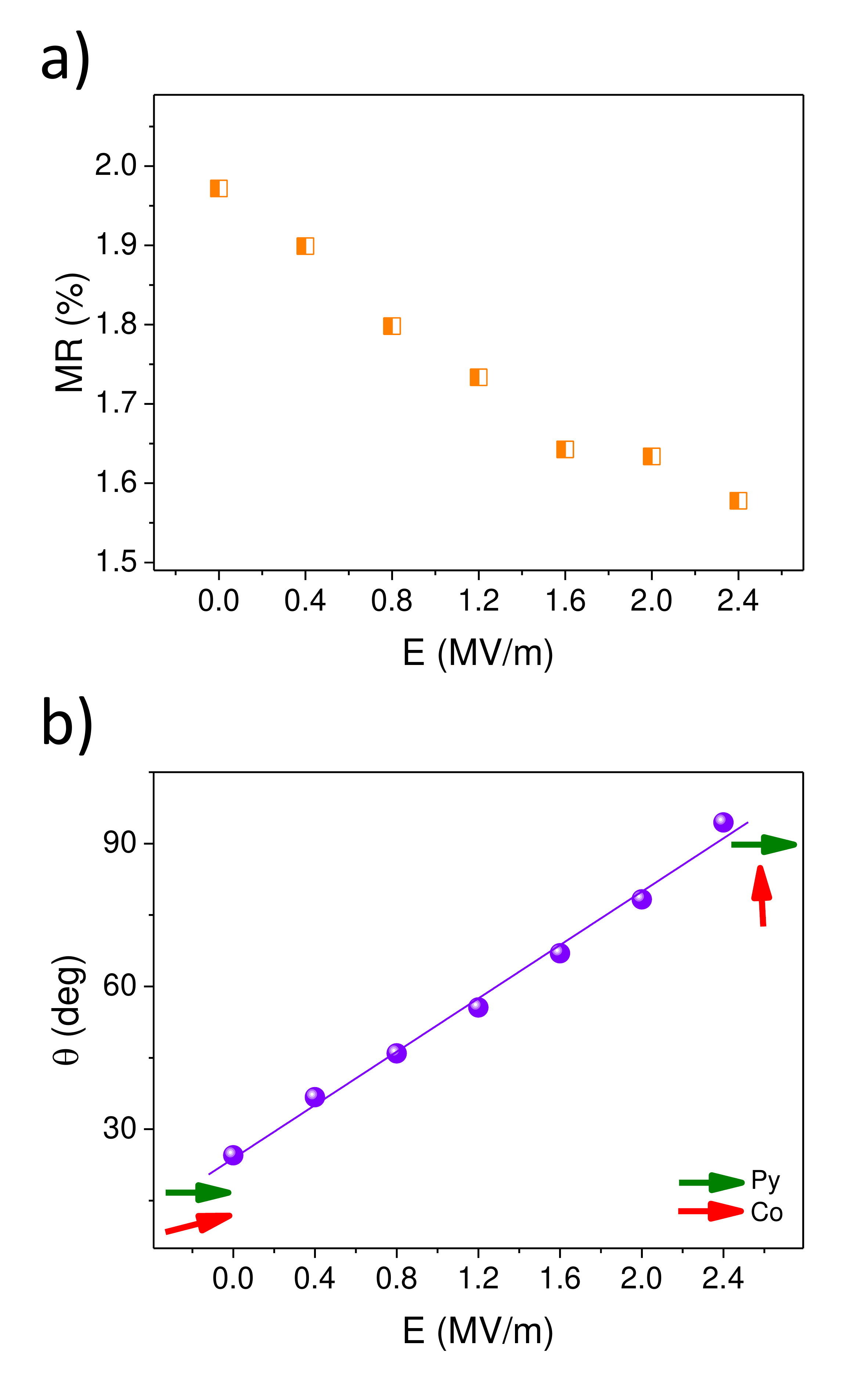}
    \caption{(a) Giant magnetoresistance ratio values extracted from Fig.~\ref{Fig2}(a) as a function of the applied electric field. (b) Angle between Co and Py net magnetizations at zero magnetic field as a function of the applied electric field. Red and green arrows denote the schematic orientation of the Co and Py magnetizations at zero magnetic field.}
    \label{Fig3}
\end{figure}  

Now we comment on the GMR ratio dependence on the applied electric field (Fig.~\ref{Fig3}(a)). The GMR ratio decrease is consistent with the strain-induced magnetic anisotropy of the Co layer, resulting in the deviation of the Co magnetization from its initial direction through a coherent magnetization rotation towards the tensile strain direction (here $y$-axis). At the MR maximum and upon increasing electric field, the angle between the magnetizations of Co and Py layers decreases (as compared to the initial close-to-antiparallel state), which therefore reduces the GMR ratio. On the other hand, at zero magnetic field, the angle between the magnetization directions of Co and Py layers increases with increasing the electrically induced strain (Fig.~\ref{Fig2}(a)). Fig.~\ref{Fig3}(b) shows the angle $\theta$ between the Co and Py net magnetizations as a function of the electric field, and calculated based on the zero magnetic field MR values of Fig.~\ref{Fig2}(a) using the relation $R(\theta) = R_P + (R_{AP} - R_P) \frac{1-cos\theta}{2} $~\cite{dieny_giant_1991} , where $R_P$ ($R_{AP}$) is the resistance in parallel (antiparallel) state. A monotonous increase of this angle suggests that the electric field induced change of the GMR ratio originates from the modification of the relative orientation of the Co and Py magnetizations. Since 
the Py layer is insensitive to strain (given its close-to-zero magnetostriction constant), we can associate this increase of the angle $\theta$ to the rotation of the Co layer magnetization only.

\begin{figure}[b]
    \centering
    \includegraphics[width=0.5\textwidth]{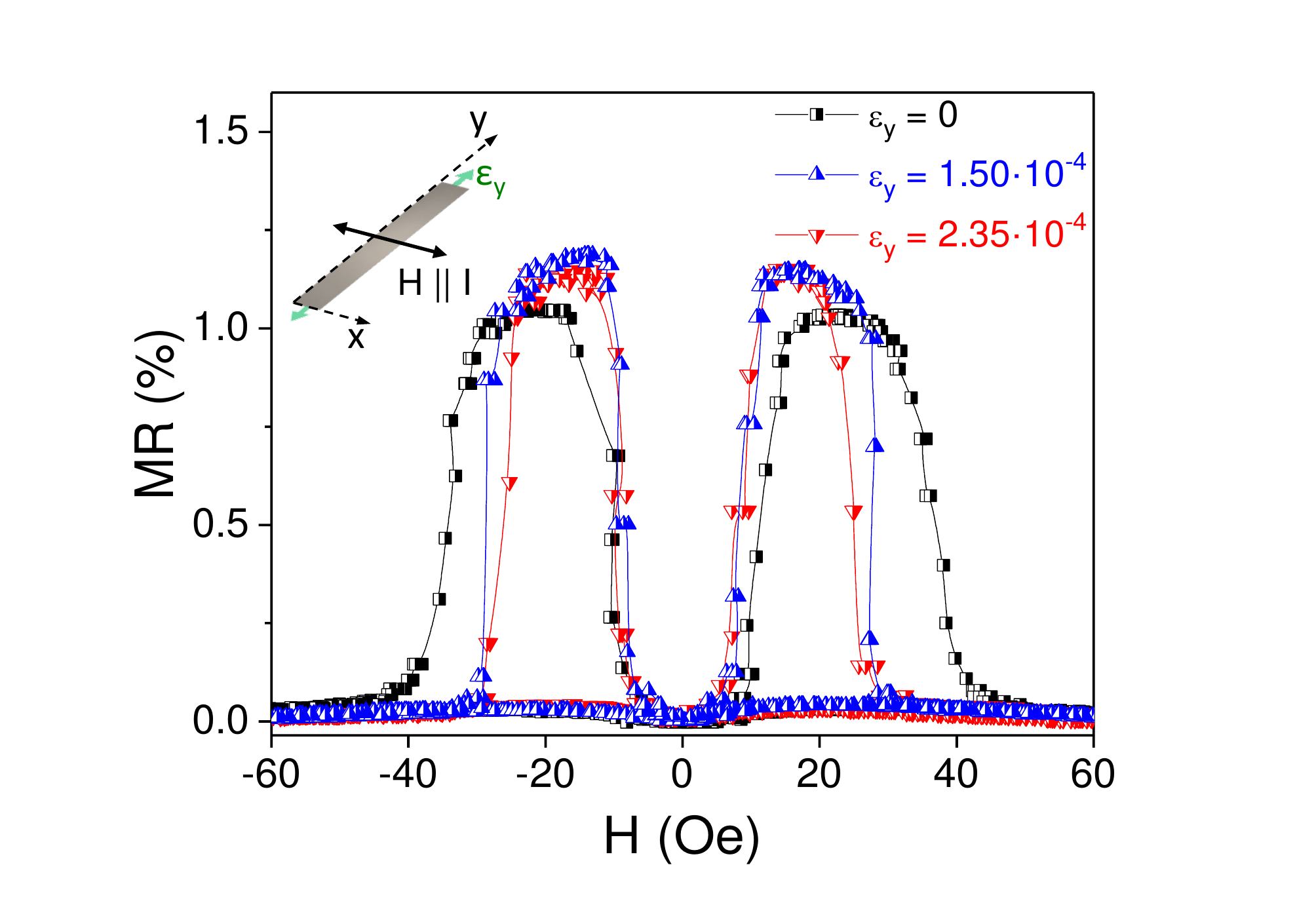}
    \caption{Magnetoresistance loops of Co(3.7)/Cu(4.5)/Py(5) stripe on Si substrate measured for the different strains generated via the mechanical bending of the substrate. The values of the strain were estimated from the curvature of the sample support (see suppl. data for more details).}
    \label{Fig4}
\end{figure}  

The strain-related origin of the observed effects is confirmed by a control experiment where the MR of the Co/Cu/Py PSV fabricated on a (100)Si substrate was measured as a function of the mechanical strain generated by bending the substrate (see details in suppl. data). Fig.~\ref{Fig4} shows the two-point MR measurements performed on the Co(3.7)/Cu(4.5)/Py(5) PSV grown on (100)Si substrate for $H\parallel I$ as a function of the tensile strain generated in the direction transverse to the applied magnetic field. The unstrained sample exhibits the MR ratio of about 1.1 $\%$ and the well-defined plateau between the switching fields $H_{sw}^{Py} \approx $ 14 Oe and $H_{sw}^{Co} \approx $ 40 Oe. The tensile uniaxial strain, generated along the $y$-direction and transverse to the MR probing direction (see inset of Fig.~\ref{Fig4}), leads to the modification of the MR properties of the PSV. The MR loops measured under strain exhibit narrower high-resistance plateaus with the decreased switching field of the Co layer whereas the switching field of the Py layer remains nearly unchanged. Similarly to the described above, this effect refers to the increase of the Co film anisotropy along the $y$-direction due to the increased magnetoelastic energy contribution introduced by the deformation of the substrate. The magnetic anisotropy of the Py layer is marginally affected by the strain due to the near-zero magnetostriction of the Py film. The observed behaviour qualitatively reproduces the results obtained for PMN-PT/Co/Cu/Py PSV described above, thus confirming the strain-related origin of the observed effects on the switching fields of the magnetostrictive Co layer.

Contrary to the case of ferroelectric PMN-PT substrate, a slight increase of the MR ratio is observed, which may be attributed to the enhanced spin-dependent electron scattering through the Cu spacer or to the modified energy landscape of the Co/Cu/Py tri-layer subjected to the complex deformation (including non-uniform bending) rather than to a simple in-plane biaxial strain.

Finally, we comment on the perspectives of the strain-controlled magnetic anisotropy and magnetotransport in magnetostrictive ferromagnetic components for applications in emerging micro- and nanoscale magnetic devices. A proper selection of the geometry and materials comprising the magnetic multilayers is a keystone of improving the efficiency of the strain-tunable magnetic elements, which may potentially be integrated in the strain-controlled spintronic sensors or memory devices~\cite{wang_strain-mediated_2017, wang_strain-mediated_2018}. Moreover, properly adjusted magnetostrictive properties of the constituent magnetic layers enable a realization of a strain-mediated switch between quantum (GMR) and classical (AMR) magnetotransport effects. Importantly, the lateral control also helps avoiding charge injection issues, common for devices with vertical electrical access, where the voltage is applied across the thickness of the magnetic element. In addition, lateral electric-field creates a room for a direct vertical access to the device for other types of excitation, e.g. light illumination or irradiation by ion beams.

As another example, when applied to spin-torque nano-oscillators, the reported effects may be used for controlling the spin polarization direction of the polarizing layer or the magnetic anisotropy of the free layer, which may broaden the dynamical range of such devices. Additionally, our findings may be utilized for controlling the dynamical coupling between the magnetic layers in magnetostrictive/non-magnetostrictive structures for efficient multilayer-based spintronic oscillators~\cite{iurchuk_stress-induced_2021}. As shown by the control experiment, the origin of strain needed for the magnetization tuning is not restricted to the converse piezoelectricity; therefore, the proposed concept can be effectively implemented in any system regardless the strain generation method. For example, recently demonstrated optically induced tuning of magnetic properties by photostriction~\cite{kundys_light_2012, iurchuk_optical_2016} may provide a new method of light-induced magnetic switching via strain-mediated mechanism.

In summary, we have demonstrated the lateral electric field induced strain-mediated control of giant magnetoresistance in PMN–PT/Co/Cu/Py pseudo spin valve. We showed that the GMR ratio and the magnetization reversal process are modified due to different magnetostrictive properties of the magnetic layers comprising the PSV. In particular, the magnetic anisotropy of the Co layer can be effectively tuned by the piezoelectric strain generated by a ferroelectric substrate, while keeping the Py layer anisotropy unaffected. This result provides an understanding of the necessity to take into account the difference in magnetostrictive properties of the magnetic films while designing strain-controllable layered magnetic systems.

\section*{Supplementary material}
See supplementary material for the details on the mechanical strain generation using convex-shaped sample supports.

\section*{Acknowledgments}
The authors gratefully acknowledge the support of the French National Research Agency via hvSTRICTSPIN ANR-13-JS 04-0008-01, Labex NIE 11-LABX-0058-NIE and ANR-10-IDEX-0002-02 research grants.

\section*{Data availability statement}
The data that support the findings of this study are available from the corresponding author upon reasonable request.

\bibliographystyle{apsrev4-1}
\bibliography{references}

\end{document}



\title{Supplementary data for "A strain-controlled magnetostrictive pseudo spin valve"}


\author{Vadym Iurchuk}
\thanks{Current affiliation: Institute of Ion Beam Physics and Materials Research, Helmholtz-Zentrum Dresden-Rossendorf, 01328 Dresden, Germany}
\email[e-mail: ]{v.iurchuk@hzdr.de}
\affiliation{Institut de Physique et Chimie des Matériaux de Strasbourg (IPCMS), UMR 7504 CNRS, Université de Strasbourg, 23 rue du Loess, 67034, Strasbourg, France}

\author{Julien Bran}
\affiliation{Institut de Physique et Chimie des Matériaux de Strasbourg (IPCMS), UMR 7504 CNRS, Université de Strasbourg, 23 rue du Loess, 67034, Strasbourg, France}

\author{Manuel Acosta}
\affiliation{Institut de Physique et Chimie des Matériaux de Strasbourg (IPCMS), UMR 7504 CNRS, Université de Strasbourg, 23 rue du Loess, 67034, Strasbourg, France}


\author{Bohdan Kundys}
\affiliation{Institut de Physique et Chimie des Matériaux de Strasbourg (IPCMS), UMR 7504 CNRS, Université de Strasbourg, 23 rue du Loess, 67034, Strasbourg, France}

\maketitle

\begin{figure}[h!]
\centering
\includegraphics[width=0.95\textwidth]{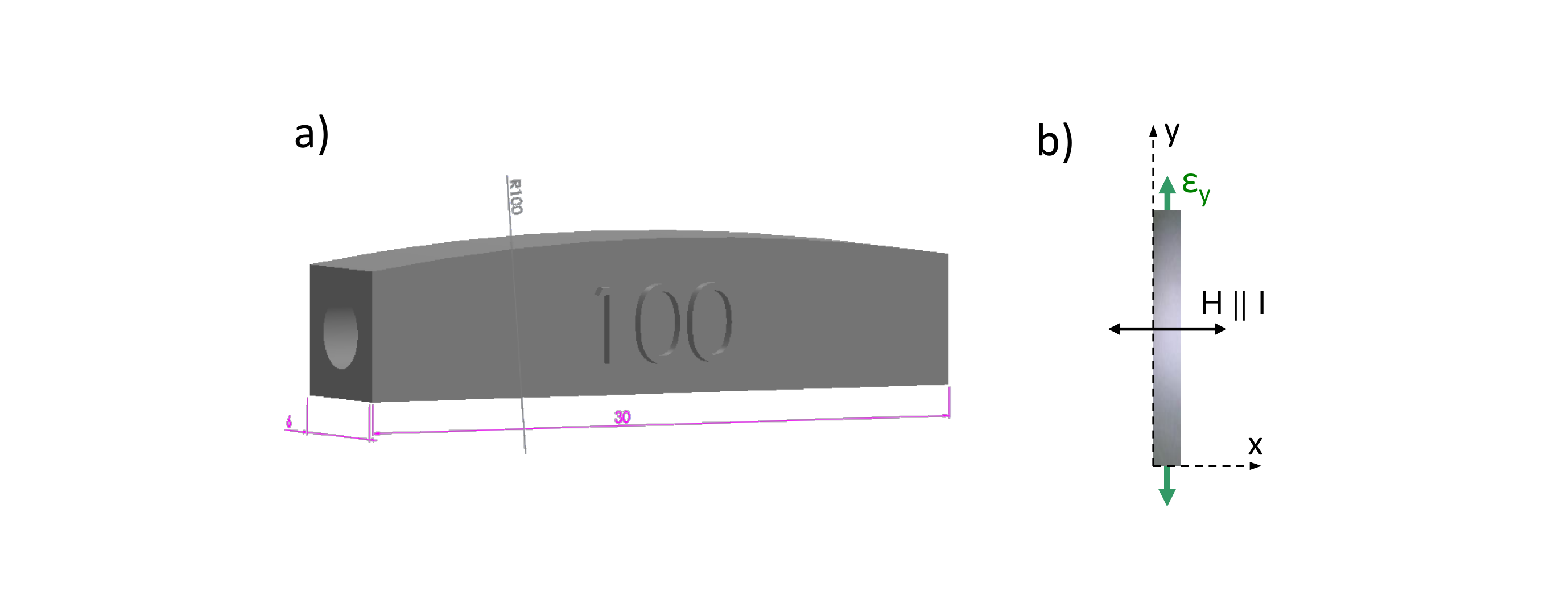}
\caption{(a) Image of the designed convex sample holder with 100 mm curvature radius. (b) Schematically sketched direction of a tensile strain generated by a convex holder.}
\label{FigS1}
\end{figure}

The $\sim$30$\times$6 mm$^2$ (100)Si platelets were used as the substrates for the Co/Cu/Py deposition. To generate a mechanical strain, polymer sample supports were 3D printed. The supports are 300$\times$6$\times$6 mm$^3$ quasi-rectangular parallelepipeds with a convex top face and different curvatures (see Fig.~\ref{FigS1}(a)). The Si/Co/Cu/Py samples are tightly attached to the top face of the support using clamps. Therefore the sample bends conformably to the shape of the support. This allows for a generation of the tensile strain along the y-direction of the sample (see Fig.~\ref{FigS1}(b)). Since the thickness of the magnetic tri-layer is negligible comparing to the Si substrate thickness and the bending angle is small, the generated tensile strain along the y-direction is considered uniform and can be calculated using relation:

\begin{equation}
    \epsilon_y = \frac{1}{\beta} \arcsin \beta - 1 
\end{equation}
with $\beta = \frac{l_0}{2r}$. Here $l_0$ is the length of the unstrained sample; $r$ is the curvature radius of the support.

The values of the generated strain for the corresponding curvatures of the support are given in Table~\ref{table1}. 
A fragility of the Si substrates does not allow reaching large values of the deformation, therefore mainly supports with 600, 500 and 400 mm curvature radii were used in this work.

\begin{table}[h!]
\centering
\caption{Generated tensile strain values for different curvature radii of the supports.}
\begin{tabular}{|c|c|} 
 \hline
 Curvature radius $r$, mm & Tensile strain $\epsilon_y$  \\
 \hline
 600 & 1.04 $\times$ 10$^{-4}$  \\ 
 \hline
 500 & 1.50 $\times$ 10$^{-4}$  \\
 \hline
 400 & 2.35 $\times$ 10$^{-4}$  \\
 \hline
 300 & 4.17 $\times$ 10$^{-4}$  \\
 \hline
\end{tabular}
\label{table1}
\end{table}